\documentstyle[12pt]{article}
\newlength{\extraspace}
\newlength{\extraspaces}
\catcode`\@=11
\def\numberbysection{\@addtoreset{equation}{section}
\def\theequation{\arabic{section}.\arabic{equation}}}
\newcommand{\be}{\begin{equation}
\addtolength{\abovedisplayskip}{\extraspaces}
\addtolength{\belowdisplayskip}{\extraspaces}
\addtolength{\abovedisplayshortskip}{\extraspace}
\addtolength{\belowdisplayshortskip}{\extraspace}}
\newcommand{\ee}{\end{equation}}
\newcommand{\ba}{\begin{eqnarray}
\addtolength{\abovedisplayskip}{\extraspaces}
\addtolength{\belowdisplayskip}{\extraspaces}
\addtolength{\abovedisplayshortskip}{\extraspace}
\addtolength{\belowdisplayshortskip}{\extraspace}}
\newcommand{\ea}{\end{eqnarray}}
\newcommand{\newsection}[1]{
\vspace{7mm}
\pagebreak[3]
\addtocounter{section}{1}
\setcounter{subsection}{0}
\setcounter{footnote}{0}
\begin{center}
{\large {\bf \thesection. #1}}
\end{center}
\nopagebreak
\medskip
\nopagebreak
\hspace{3mm}}
\newcommand{\nonu}{\nonumber \\[.5mm]}
\newcommand{\A}{&\!\!\!}

\newcommand{\tr}{\, {\rm tr}}
\newcommand{\e}{\, {\rm e}}

\begin{document}
\addtolength{\baselineskip}{.7mm}
\thispagestyle{empty}
\begin{flushright}
OCHA--PP--143 \\
STUPP--99--158 \\
{\tt hep-th/9910192} \\ 
October, 1999
\end{flushright}
\vspace{7mm}
\begin{center}
{\Large{\bf Local Symmetries in the AdS${}_7$/CFT${}_6$ 
Correspondence}} \\[20mm] 
{\sc Madoka Nishimura}
\\[3mm]
{\it Department of Physics, Ochanomizu University \\
2-1-1, Otsuka, Bunkyo-ku, Tokyo 112-0012, Japan} \\[3mm]
and \\[3mm]
{\sc Yoshiaki Tanii}\footnote{
\tt e-mail: tanii@post.saitama-u.ac.jp} \\[3mm]
{\it Physics Department, Faculty of Science \\
Saitama University, Urawa, Saitama 338-8570, Japan} \\[20mm]

{\bf Abstract}\\[7mm]
{\parbox{13cm}{\hspace{5mm}
It is shown that local symmetry transformations of the maximal AdS 
supergravity in seven-dimensional anti de Sitter space induce 
those of the $N=(2,0)$ conformal supergravity on the six-dimensional 
boundary at infinity. Boundary values of the AdS supergravity fields 
form a supermultiplet of the conformal supergravity. 
}}
\end{center}
%
%
\newsection{Introduction}
In the AdS/CFT correspondence \cite{MAL,GKP,WITTEN} (For a review 
see, e.g., ref.\ \cite{AGMOO}.) a supergravity in ($d+1$)-dimensional 
anti de Sitter space is dual to a conformal field theory on 
$d$-dimensional boundary at infinity. Boundary values of 
supergravity fields play a role of sources for operators of the 
conformal field theory \cite{GKP,WITTEN}. It was noted in 
refs.\ \cite{FFZ,LT} that these boundary fields should form 
supermultiplets of a conformal supergravity in $d$ dimensions. 
Conformal supergravities are theories which have Weyl and super 
Weyl symmetries in addition to local symmetries of the Poincar\'e 
supergravities. 
\par
In refs.\ \cite{NT,NT2} we explicitly showed such a relation 
between AdS supergravities and conformal supergravities in the 
case of three-dimensional anti de Sitter space. It was shown 
that local symmetry transformations of the bulk AdS supergravities 
induce local transformations of the boundary fields, which 
coincide with those of two-dimensional conformal supergravities. 
In particular, Weyl and super Weyl transformations on the boundary 
are generated from general coordinate and super transformations 
in the bulk. Thus, the boundary fields form a supermultiplet of the 
conformal supergravities. 
A relation between local symmetries in the bulk and those 
on the boundary is also discussed in ref.\ \cite{BAUTIER}. 
\par
The purpose of this paper is to study a similar relation between 
an AdS supergravity and a conformal supergravity in the case of 
seven-dimensional anti de Sitter space. It is shown that local 
transformations of the maximal AdS supergravity in seven 
dimensions \cite{PPvN} induce those of the $N=(2,0)$ conformal 
supergravity in six dimensions \cite{BSvP} on the boundary. 
The seven-dimensional maximal supergravity appears in a 
compactification of the M-theory on AdS${}_7$ $\times$ S${}^4$ 
\cite{NVvN}, which corresponds to a configuration of $N$ M5-branes 
in the large $N$ limit \cite{MAL}. A new feature in this case is 
the presence of third rank antisymmetric tensor fields in the 
seven-dimensional supergravity. They satisfy a so-called 
``self-duality in odd dimensions'' \cite{TPvN} by field equations. 
We show that they become self-dual antisymmetric tensor fields 
of the six-dimensional conformal supergravity on the boundary. 
\par
As in refs.\ \cite{NT,NT2} we partially fix the gauge for local 
symmetries in the bulk. Our gauge choice is a sort of axial gauge, 
in which the direction normal to the boundary is distinguished. 
We first obtain boundary behaviors of all the fields and the 
gauge transformation parameters after the gauge fixing by field 
equations and residual symmetry equations. Substituting 
them into the local symmetry transformations we obtain the 
conformal supergravity transformations on the boundary. 
\par
%
%
\newsection{Maximal AdS supergravity in seven dimensions}
The field content of the maximal AdS supergravity in seven 
dimensions \cite{PPvN} is a vielbein $e_M{}^A$, 
Rarita-Schwinger fields $\psi_M^\alpha$, real third rank 
antisymmetric tensor fields $S_{MNP,I}$, SO(5)$_g$ vector fields 
$B_{M,I}{}^J$, spin ${1 \over 2}$ spinor fields $\lambda_i^\alpha$ 
and scalar fields $\Pi_I{}^i$. We denote seven-dimensional world 
indices as $M, N, \cdots = 0, 1, \cdots, 6$ and local Lorentz 
indices as $A, B, \cdots = 0, 1, \cdots, 6$. 
The indices $I, J, \cdots = 1, \cdots, 5$ are vector indices 
of SO(5)${}_g$, while $i, j, \cdots = 1, \cdots, 5$ and 
$\alpha, \beta, \cdots = 1, \cdots, 4$ are vector and spinor 
indices of SO(5)${}_c$ respectively. 
SO(5)${}_g$ and SO(5)${}_c$ are local symmetries of the theory, 
which will be discussed later. 
The flat metric is $\eta_{AB} = {\rm diag}(-1, +1, \cdots, +1)$ 
and the totally antisymmetric tensor $\epsilon^{M_1 \cdots M_7}$ 
is chosen as $\epsilon^{0123456} = +1$. 
We need two kinds of gamma matrices: $8 \times 8$ matrices 
$\gamma^A$ for the seven-dimensional Lorentz group SO(1,6) and 
$4 \times 4$ matrices $\tau^i$ for SO(5)${}_c$ satisfying 
$\{ \gamma^A, \gamma^B \} = 2 \eta^{AB}$, 
$\{ \tau^i, \tau^j \} = 2 \delta^{ij}$. 
$\gamma$'s and $\tau$'s with multiple indices are antisymmetrized 
products of gamma matrices with unit strength. In particular, 
we have $\gamma^{A_1 \cdots A_7} = - \epsilon^{A_1 \cdots A_7}$. 
The Dirac conjugate of a spinor $\psi$ 
is defined as $\bar\psi = \psi^\dagger i \gamma^0$. 
The spinor fields satisfy a symplectic Majorana condition 
$\psi^\alpha = \Omega^{\alpha\beta} C \bar\psi_\beta^T$, 
where $C = C^T$ and $\Omega = - \Omega^T$ are charge conjugation 
matrices of SO(1,6) and SO(5)${}_c$ satisfying 
\be
C^{-1} \gamma^A C = - (\gamma^A)^T, \qquad
\Omega^{-1} \tau^i \Omega = (\tau^i)^T. 
\ee
The spinor fields $\lambda_i^\alpha$ also satisfy the 
SO(5)${}_c$ irreducibility condition 
$(\tau^i)^\alpha{}_\beta \lambda_i^\beta = 0$. 
The scalar fields $\Pi_I{}^i$ satisfy $\det \Pi_I{}^i = 1$, i.e., 
$\Pi_I{}^i \in$ SL(5, {\bf R}). By the local SO(5)${}_c$ symmetry 
physical degrees of freedom of the scalar fields parametrize 
a coset space SL(5, {\bf R})/SO(5)${}_c$. 
\par
The Lagrangian is given by 
\ba
{\cal L} \A = \A 
-{1 \over 2} e R + 4 m^2 e \left( T^2 - 2 T_{ij} T^{ij} \right)
- {1 \over 2} e P_{Mij} P^{Mij} 
- {1 \over 4} e \left( F_{MN}{}^{IJ} \Pi_I{}^i \Pi_J{}^j 
\right)^2 \nonu
\A \A + 8 m^2 e \left( \Pi^{-1}{}_i{}^I S_{MNP,I} \right)^2 
+ {1 \over 12} m \epsilon^{MNPQRST} S_{MNP,I} 
F_{QRST}{}^I \nonu
\A \A - {1 \over 2} e \bar\psi_M \gamma^{MNP} D_N \psi_P 
- {1 \over 2} e \bar\lambda^i \gamma^M D_M \lambda_i 
+ {1 \over 2} m e T \bar\psi_M \gamma^{MN} \psi_N \nonu
\A \A - {1 \over 2} m e \left( 8 T^{ij} - T \delta^{ij} \right) 
\bar\lambda_i \lambda_j 
+ 2 m e T^{ij} \bar\lambda_i \tau_j \gamma^M \psi_M 
+ {1 \over 2} e \bar\psi_M \gamma^N \gamma^M \tau^i 
\lambda^j P_{Nij} \nonu
\A \A + {1 \over 16} e \bar\psi_M \left( \gamma^{MNPQ} 
- 2 g^{MN} g^{PQ} \right) \tau_{ij} \psi_Q 
F_{NP}{}^{IJ} \Pi_I{}^i \Pi_J{}^j \nonu
\A \A + {1 \over 4} e \bar\psi_M \gamma^{NP} \gamma^M 
\tau_i \lambda_j F_{NP}{}^{IJ} \Pi_I{}^i \Pi_J{}^j 
+ {1 \over 32} e \bar\lambda_i \tau^j \tau_{kl} \tau^i 
\gamma^{MN} \lambda_j F_{MN}{}^{IJ} \Pi_I{}^k \Pi_J{}^l \nonu
\A \A + { 1 \over 2 \sqrt{3}} i m e \bar\psi_M \left( 
\gamma^{MNPQR} + 6 g^{MN} \gamma^P g^{QR} \right) \tau^i \psi_R 
\Pi^{-1}{}_i{}^I S_{NPQ,I} \nonu
\A \A - {1 \over \sqrt{3}} i m e \bar\psi_M \left( 
\gamma^{MNPQ} - 3 g^{MN} \gamma^{PQ} \right) 
\lambda^i \Pi^{-1}{}_i{}^I S_{NPQ,I} \nonu
\A \A - {1 \over 2 \sqrt{3}} i m e \bar\lambda^i \gamma^{MNP} \tau^j 
\lambda_i \Pi^{-1}{}_j{}^I S_{MNP,I} 
+ {1 \over 32m} \Omega_5[B] - {1 \over 64m} \Omega_3[B] \nonu
\A \A + {1 \over 16 \sqrt{3}} \epsilon^{MNPQRST} 
\epsilon_{IJ_1 J_2 J_3 J_4} \delta^{IJ} S_{MNP}{}^I 
F_{QR}{}^{J_1 J_2} F_{ST}{}^{J_3 J_4} + \cdots, 
\label{lag}
\ea
where we have put the gravitational constant as $8\pi G = 1$, 
$m$ is a positive constant and the dots denote four-fermi terms. 
The quantities in eq.\ (\ref{lag}) are defined as follows. 
{}From the scalar fields we define 
\ba
P_{M(ij)} + Q_{M[ij]} \A = \A 
(\Pi^{-1})_i{}^I \left( \partial_M \Pi_{Ij} + 
8m B_{MI}{}^J \Pi_{Jj} \right), \nonu
T_{ij} \A = \A (\Pi^{-1})_i{}^I (\Pi^{-1})_j{}^J \delta_{IJ}, \qquad 
T = T_{ij} \delta^{ij}, 
\ea
where $(ij)$ and $[ij]$ are symmetric and antisymmetric parts. 
The field strengths of the vector fields and the antisymmetric 
tensor fields are 
\ba
F_{MN,I}{}^J \A = \A 
\partial_M B_{N,I}{}^J + 8 m B_{M,I}{}^K B_{N,K}{}^J 
- (M \leftrightarrow N), \nonu
F_{MNPQ,I} \A = \A 4 D_{[M} S_{NPQ]I}. 
\ea
The SO(5)${}_g$ gauge coupling constant is $8m$. 
The covariant derivative $D_M$ contains the SO(5)${}_g$ gauge field 
$B_{MI}{}^J$ and the composite SO(5)${}_c$ gauge field $Q_{M[ij]}$ 
as well as the spin connection $\omega_M{}^{AB}$, e.g., 
\be
D_M \lambda_i^\alpha 
= \left( \partial_M + {1 \over 4} \omega_M{}^{AB} 
\gamma_{AB} \right) \lambda_i^\alpha 
+ Q_{Mi}{}^j \lambda_j^\alpha + {1 \over 4} Q_M{}^{jk} 
(\tau_{jk})^\alpha{}_\beta  \lambda_i^\beta. 
\ee
Finally, $\Omega_3[B]$ and $\Omega_5[B]$ are Chern-Simons terms 
satisfying in the differential form language 
\be
d \Omega_5[B] = 8 \tr( F^4 ), \qquad
d \Omega_3[B] = 8 \tr( F^2 ) \tr( F^2 ). 
\ee
\par
The Lagrangian (\ref{lag}) is invariant under general coordinate, 
local Lorentz, local SO(5)${}_g$, local SO(5)${}_c$ and local 
super transformations up to total derivative terms. 
Note that there are neither Weyl nor super Weyl 
symmetries in the bulk of seven-dimensional spacetime. 
The bosonic transformations are 
\ba
\delta e_M{}^A 
\A = \A \xi^N \partial_N e_M{}^A + \partial_M \xi^N e_N{}^A 
- \lambda^A{}_B e_M{}^B, \nonu
\delta \psi_M 
\A = \A \xi^N \partial_N \psi_M + \partial_M \xi^N \psi_N
- {1 \over 4} \lambda^{AB} \gamma_{AB} \psi_M
- {1 \over 4} v^{ij} \tau_{ij} \psi_M, \nonu
\delta S_{MNP,I} 
\A = \A \xi^Q \partial_Q S_{MNP,I} 
+ 3 \partial_{[P} \xi^Q S_{MN]Q,I} 
- \theta_I{}^J S_{MNP,J}, \nonu
\delta B_M{}^{IJ}
\A = \A \xi^N \partial_N B_M{}^{IJ} 
+ \partial_M \xi^N B_N{}^{IJ} + D_M \theta^{IJ}, \nonu
\delta \lambda_i 
\A = \A \xi^M \partial_M \lambda_i
- {1 \over 4} \lambda^{AB} \gamma_{AB} \lambda_i
- v_i{}^j \lambda_j - {1 \over 4} v^{jk} \tau_{jk} \lambda_i, \nonu
\delta \Pi_I{}^i 
\A = \A \xi^M \partial_M \Pi_I{}^i
- \theta_I{}^J \Pi_J{}^i - v^i{}_j \Pi_I{}^j, 
\label{bsym}
\ea
where $\xi^M(x)$, $\lambda^{AB}(x) = - \lambda^{BA}(x)$, 
$\theta^{IJ}(x) = - \theta^{JI}(x)$ and $v^{ij}(x) = -v^{ji}(x)$ 
are transformation parameters of general coordinate, local Lorentz, 
SO(5)${}_g$ and SO(5)${}_c$ transformations respectively. 
Note that there is no antisymmetric tensor gauge symmetry 
for $S_{MNP,I}$, which satisfy the ``self-duality in odd 
dimensions'' \cite{TPvN} by field equation. 
\par
The local supertransformations are 
\ba
\delta e_M{}^A 
\A = \A {1 \over 2} \bar\epsilon \gamma^A \psi_M, \nonu
\delta \psi_M 
\A = \A D_M \epsilon + {1 \over 5} m T \gamma_M \epsilon 
- {1 \over 40} ( \gamma_M{}^{NP} - 8 \delta_M^N \gamma^P ) 
\tau_{ij} \epsilon F_{NP}{}^{IJ} \Pi_I{}^i \Pi_J{}^j \nonu
\A \A + {2 \over 5 \sqrt{3}} i m \left( \gamma_M{}^{NPQ} 
- {9 \over 2} \delta_M^N \gamma^{PQ} \right) \tau^i \epsilon 
\Pi^{-1}{}_i{}^I S_{NPQ,I} + \cdots, \nonu
\delta S_{MNP,I}
\A = \A {\sqrt{3} \over 12} i \left( 
3 \bar\epsilon \gamma_{[MN} \tau^i \psi_{P]} 
- \bar\epsilon \gamma_{MNP} \lambda^i \right) \Pi^{-1}{}_{iI} \nonu
\A \A - {\sqrt{3} \over 32 m} i \Pi_I{}^i \left( 
2 \bar\epsilon \tau_{ijk} \psi_{[M} 
+ \bar\epsilon \gamma_{[M} \tau^l \tau_{ijk} \lambda_l \right)
F_{NP]}{}^{JK} \Pi_J{}^j \Pi_K{}^k \nonu
\A \A - {\sqrt{3} \over 16 m} i D_{[M} \left( 
2 \bar\epsilon \gamma_N \tau^i \psi_{P]} 
+ \bar\epsilon \gamma_{NP]} \lambda^i \right) 
\Pi^{-1}{}_{iI}, \nonu
\delta B_M{}^{IJ}
\A = \A \left( {1 \over 4} \bar\epsilon \tau^{ij} \psi_M 
+ {1 \over 8} \bar\epsilon \gamma_M \tau^k \tau^{ij} \lambda_k 
\right) \Pi^{-1}{}_i{}^I \Pi^{-1}{}_j{}^J, \nonu
\delta \lambda_i 
\A = \A 2 m \left( T_{ij} - {1 \over 5} \delta_{ij} T 
\right) \tau^j \epsilon 
+ {1 \over 16} \gamma^{MN} \left( \tau_{kl} \tau_i 
- {1 \over 5} \tau_i \tau_{kl} \right) \epsilon 
F_{MN}{}^{IJ} \Pi_I{}^k \Pi_J{}^l \nonu
\A \A + {1 \over 5 \sqrt{3}} i m \gamma^{MNP} \left( 
\tau_i{}^j - 4 \delta_i^j \right) \epsilon \Pi^{-1}{}_j{}^I S_{MNP,I} 
+ {1 \over 2} \gamma^M \tau^j \epsilon P_{Mij} + \cdots, \nonu
\delta \Pi_I{}^i 
\A = \A {1 \over 4} \Pi_I{}^j \left(
\bar\epsilon \tau_j \lambda^i + \bar\epsilon \tau^i \lambda_j 
\right), 
\label{fsym}
\ea
where the dots denote three-fermi terms, which we ignore in 
the following. The transformation parameter $\epsilon^\alpha(x)$ 
satisfies the symplectic Majorana condition. 
%
%
\newsection{Boundary behaviors of the fields}
We partially fix the gauge for the local symmetries (\ref{bsym}) 
and (\ref{fsym}). The seven-dimensional AdS space is represented as 
a region $r \equiv x^6 > 0$ in ${\bf R}^7$ with coordinates $x^M$. 
The boundary of the AdS space corresponds to a hyperplane $r = 0$ 
and a point $r = \infty$. We choose the gauge fixing condition as 
\ba
e_r{}^6 \A = \A {1 \over 2mr}, \qquad
e_r{}^a = 0, \qquad
e_\mu{}^6 = 0, \nonu
\psi_r \A = \A 0, \qquad
B_r{}^{IJ} = 0, \qquad
\Pi^T = \Pi, 
\label{gcond}
\ea
where $\mu, \nu, \cdots = 0, \cdots, 5$ and 
$a, b, \cdots = 0, \cdots, 5$ are six-dimensional world indices 
and local Lorentz indices respectively. 
The metric in this gauge has a form 
\be
dx^M dx^N g_{MN} 
= {1 \over (2mr)^2} \left( 
dr^2 + dx^\mu dx^\nu \hat g_{\mu\nu} \right). 
\ee
The SO(6,2) invariant AdS metric corresponds to the case 
$\hat g_{\mu\nu} = \eta_{\mu\nu}$ but we consider the general 
$\hat g_{\mu\nu}$. We define $\hat e_\mu{}^a$ by 
$\hat g_{\mu\nu} = \hat e_\mu{}^a \hat e_\nu{}^b \eta_{ab}$. 
An SO(6,2) invariant field configuration 
\be
\hat g_{\mu\nu} = \eta_{\mu\nu}, \qquad 
\Pi_I{}^i = \delta_I^i, \qquad
\mbox{other fields = 0}. 
\label{background}
\ee
is a solution of field equations derived from the Lagrangian 
(\ref{lag}). 
\par
Let us obtain asymptotic behaviors of the fields near the boundary 
$r = 0$. The boundary conditions are chosen such that 
they are consistent with these boundary behaviors. 
We assume that the vielbein $e_\mu{}^a$ behaves as $r^{-1}$ 
just as in the SO(6,2) invariant case (\ref{background}). 
Boundary behaviors of other fields are determined by their 
field equations. For the scalar fields $\Pi_I^i$ we first expand 
them around the background (\ref{background}) as 
\be
\Pi_I{}^i = ( \e^\phi )_I{}^i 
= \delta_I^j \left( \delta_j^i + \phi_j{}^i 
+ {1 \over 2} \phi_j{}^k \phi_k{}^i + {\cal O}(\phi^3) \right), 
\ee
where $\phi_{ij} = \phi_{ji}$ and $\phi_i{}^i = 0$. 
Near the boundary $r = 0$ the linearized field equation for 
the scalar fields in the background (\ref{background}) is 
\be
r^7 \partial_r \left( r^{-5} \partial_r \phi \right) 
+ 8 \phi = 0. 
\ee
Assuming $\phi \sim r^s$ for $r \rightarrow 0$, we obtain 
two independent solutions: $\phi \sim r^2$ and $\phi \sim r^4$. 
The solution regular in the bulk is a linear combination 
of these two solutions and its boundary behavior is determined 
by a solution which becomes larger near the boundary, i.e., 
$\phi \sim r^2$. The same reasoning can be applied to other 
fields discussed below. 
For the vector fields $B_M$ the linearized field equation is 
\be 
\partial_r \left( r^{-3} \partial_r B_\mu \right) = 0. 
\ee
We find two solutions: $B_\mu \sim r^0$ and $B_\mu \sim r^4$. 
\par
For the antisymmetric tensor fields $S_{MNP}$ the linearized field 
equation is 
\be
{1 \over 6} \epsilon^{MNPQ_1 \cdots Q_4} F_{Q_1 \cdots Q_4} 
+ 16 m e S^{MNP} = 0. 
\label{anti}
\ee
By the $(MNP)=(r\mu\nu)$ components of this equation 
the components $S_{r\mu\nu}$ can be expressed by using 
$S_{\mu\nu\rho}$ as 
\be
S_{r\mu\nu} = - {r \over 12} 
\epsilon_{\mu\nu\rho_1 \cdots \rho_4} 
\eta^{\rho_1\sigma_1} \cdots \eta^{\rho_4\sigma_4} 
\partial_{[\sigma_1} S_{\sigma_2 \sigma_3 \sigma_4]}. 
\ee
Therefore they are not independent degrees of freedom and behave 
as $S_{r\mu\nu} \sim r^{s+1}$ when $S_{\mu\nu\rho} \sim r^s$. 
The $(MNP)=(\mu\nu\rho)$ components of the field equation 
(\ref{anti}) become 
\be
\partial_r S_{\mu\nu\rho} - {2 \over r} \tilde S_{\mu\nu\rho} 
- 3 \partial_{[\mu} S_{\nu\rho]r} = 0, 
\ee
where the dual of $S_{\mu\nu\rho}$ is defined as 
\be
\tilde S_{\mu\nu\rho} 
= {1 \over 3!} \hat e 
\epsilon_{\mu\nu\rho\sigma_1\sigma_2\sigma_3} 
\hat g^{\sigma_1\lambda_1} \hat g^{\sigma_2\lambda_2} 
\hat g^{\sigma_3\lambda_3} S_{\lambda_1\lambda_2\lambda_3}. 
\ee
To solve this equation we introduce self-dual and anti self-dual 
parts of $S_{\mu\nu\rho}$ as $S^{(\pm)} = {1 \over 2} \left( 
S_{\mu\nu\rho} \pm \tilde S_{\mu\nu\rho} \right)$. 
Then we find two solutions: $S^{(+)}_{\mu\nu\rho} \sim r^0$, 
$S^{(-)}_{\mu\nu\rho} \sim r^{-2}$ and 
$S^{(+)}_{\mu\nu\rho} \sim r^2$, $S^{(-)}_{\mu\nu\rho} \sim r^4$. 
\par
For spinor fields $\lambda$ the linearized field equation is 
\be
\partial_r \lambda - {3 \over 2r} ( 2 - \gamma_6) \lambda 
+ \gamma_6 \hat\gamma^\mu \partial_\mu \lambda = 0. 
\ee
We define the projections 
$\lambda_\pm = {1 \over 2} \left( 1 \pm \gamma_6 \right) \lambda$. 
Note that $\gamma_6 = - \gamma^0 \gamma^1 \cdots \gamma^5$ 
is the six-dimensional chirality matrix and $\lambda_+$, 
$\lambda_-$ are Weyl spinors. We obtain two solutions: 
$\lambda_+ \sim r^{3 \over 2}$, $\lambda_- \sim r^{5 \over 2}$ and 
$\lambda_+ \sim r^{11 \over 2}$, $\lambda_- \sim r^{9 \over 2}$. 
For the Rarita-Schwinger field $\psi_M$ the linearized field 
equation is 
\be
\hat\gamma^{\mu\nu} \left( \partial_r \psi_\nu 
- {1 \over 2r} ( 4 - 5 \gamma_6 ) \psi_\nu \right) 
- \gamma_6 \hat\gamma^{\mu\nu\rho} \partial_\nu \psi_\rho = 0. 
\ee
We find two solutions: 
$\psi_{\mu+} \sim r^{-{1 \over 2}}$, 
$\psi_{\mu-} \sim r^{1 \over 2}$ and 
$\psi_{\mu+} \sim r^{11 \over 2}$, 
$\psi_{\mu-} \sim r^{9 \over 2}$. 
\par
Knowing these boundary behaviors of the fields we impose Dirichlet 
type boundary conditions as 
\ba
e_\mu{}^a \A \sim \A (2mr)^{-1} e_{0\mu}{}^a, \qquad
\psi_{\mu+} \sim (2mr)^{-{1 \over 2}} \psi_{0\mu+}, \qquad
S_{\mu\nu\rho}^{(-)} \sim r^{-2} S_{0\mu\nu\rho}^{(-)}, \nonu
B_\mu \A \sim \A B_{0\mu}, \qquad
\lambda_+ \sim (2mr)^{3 \over 2} \lambda_{0+}, \qquad
\phi \sim (2mr)^2 \phi_0, 
\label{bbfield}
\ea
where $\Phi_0  = (e_{0\mu}{}^a, \psi_{0\mu+}, 
S^{(-)}_{0\mu\nu\rho}, B_{0\mu}, \lambda_{0+}, \phi_0)$ 
are arbitrary functions on the boundary. 
Note that we imposed the boundary conditions on only half 
of the components for $\psi_\mu$, $S_{\mu\nu\rho}$ and $\lambda$ 
since their field equations are first order \cite{HS,AF}. 
Other components of the fields are expressed by the functions 
$\Phi_0$ by using the field equations. 
The fields $\Phi_0$ coincide with the field content of the 
(2,0) conformal supergravity in six dimensions \cite{BSvP}. 
The precise relation between $\Phi_0$ and the fields of the 
conformal supergravity will be given below. 
\par
%
%
\newsection{Local symmetries on the boundary}
We now study how the fields on the boundary $\Phi_0$ transform 
under residual symmetry transformations after the gauge fixing. 
We first obtain the residual symmetries, which preserve the gauge 
conditions (\ref{gcond}). By solving differential equations 
obtained by taking variations of the gauge conditions (\ref{gcond}) 
under the transformations (\ref{bsym}), (\ref{fsym}) 
we find parameters of the residual symmetry 
transformations near the boundary $r = 0$ as 
\ba
\xi^r \A = \A -  r \Lambda_0, \nonu
\xi^\mu \A = \A \xi_0^\mu + {\cal O}(r^2), \nonu
\lambda_{a6} \A = \A {\cal O}(r), \nonu
\lambda_{ab} \A = \A \lambda_{0ab} + {\cal O}(r^2), \nonu
\theta^{IJ} \A = \A \theta_0{}^{IJ} + {\cal O}(r^2), \nonu
v^{ij} \A = \A \theta_0{}^{ij} + {\cal O}(r^2), \nonu
\epsilon_\pm \A = \A (2mr)^{\mp{1 \over 2}} \left[ 
\epsilon_{0\pm} + {\cal O}(r^2) \right], 
\label{bbparam}
\ea
where $\Lambda_0$, $\xi_0^\mu$, $\lambda_{0ab}$, 
$\theta_0^{ij} = \theta_0{}^{IJ} \delta_I^i \delta_J^j$ and 
$\epsilon_{0\pm}$ are arbitrary functions of $x^\mu$ 
($\mu=0, \cdots, 5$). Order ${\cal O}(r)$ and ${\cal O}(r^2)$ 
terms are non-local functionals of these functions and the 
fields $\Phi_0$. 
\par
Substituting eqs.\ (\ref{bbfield}), (\ref{bbparam}) into 
eq.\ (\ref{bsym}) and taking the limit $r \rightarrow 0$ we find 
the bosonic transformations on the boundary as 
\ba
\delta e_{0\mu}{}^a 
\A = \A \Lambda_0 e_{0\mu}{}^a 
+ \xi_0^\nu \partial_\nu e_{0\mu}{}^a 
+ \partial_\mu \xi_0^\nu e_{0\nu}{}^a 
- \lambda_0{}^a{}_b e_{0\mu}{}^b, \nonu
\delta \psi_{0\mu+} 
\A = \A {1 \over 2} \Lambda_0 \psi_{0\mu+} 
+ \xi_0^\nu \partial_\nu \psi_{0\mu+} 
+ \partial_\mu \xi_0^\nu \psi_{0\nu+} 
- {1 \over 4} \lambda_0{}^{ab} \gamma_{ab} \psi_{0\mu+} 
- {1 \over 4} \theta_0{}^{ij} \tau_{ij} \psi_{0\mu+}, \nonu
\delta S^{(-)}_{0\mu\nu\rho,I} 
\A = \A 2 \Lambda_0 S^{(-)}_{0\mu\nu\rho,I} 
+ \xi_0^\sigma \partial_\sigma S^{(-)}_{0\mu\nu\rho,I} 
+ 3 \partial_{[\rho} \xi_0^\sigma S^{(-)}_{0\mu\nu]\sigma,I} 
- \theta_{0I}{}^J S^{(-)}_{0\mu\nu\rho,J}, \nonu
\delta B_{0\mu}{}^{IJ} 
\A = \A \xi_0^\nu \partial_\nu B_{0\mu}{}^{IJ} 
+ \partial_\mu \xi_0^\nu B_{0\nu}{}^{IJ} 
+ D_{0\mu} \theta_0{}^{IJ}, \nonu
\delta \lambda_{0i+} 
\A = \A - {3 \over 2} \Lambda_0 \lambda_{0i+} 
+ \xi_0^\nu \partial_\nu \lambda_{0i+} 
- {1 \over 4} \lambda_0{}^{ab} \gamma_{ab} \lambda_{0i+} 
- \theta_{0i}{}^j \lambda_{0j+} 
- {1 \over 4} \theta_0{}^{jk} \tau_{jk} \lambda_{0i+}, \nonu
\delta \phi_{0ij} 
\A = \A - 2 \Lambda_0 \phi_{0ij} 
+ \xi_0^\nu \partial_\nu \phi_{0ij} 
- \theta_{0i}{}^k \phi_{0kj} 
- \theta_{0j}{}^k \phi_{0ik}. 
\label{6dbsym}
\ea
The transformations with the parameters $\xi_0^\mu$, $\Lambda_0$, 
$\lambda_0^{ab}$ and $\theta_0^{ij}$ represent general coordinate, 
Weyl, local Lorentz and SO(5) gauge transformations in six 
dimensions respectively. In particular, seven-dimensional general 
coordinate transformation in the direction $M=r$ became 
six-dimensional Weyl transformation. Weights of the Weyl 
transformation are determined by the powers of $r$ appearing in 
the boundary behaviors of the fields (\ref{bbfield}). 
The weights given in eq.\ (\ref{6dbsym}) are consistent with 
those of the conformal supergravity \cite{BSvP}. 
\par
On the other hand, substituting eqs.\ (\ref{bbfield}), 
(\ref{bbparam}) into eq.\ (\ref{fsym}) and taking the limit 
$r \rightarrow 0$ we find the fermionic transformations 
on the boundary as 
\ba
\delta e_{0\mu}{}^a 
\A = \A {1 \over 2} \bar\epsilon_{0+} \gamma^a \psi_{0\mu+}, \nonu
\delta \psi_{0\mu+} 
\A = \A D_{0\mu} \epsilon_{0+} 
- {1 \over 2 \sqrt{3}} i m \tau^i \gamma_0^{\nu\rho\sigma} 
\gamma_{0\mu} \epsilon_{0+} S^{(-)}_{0\nu\rho\sigma,i} 
+ 2m \gamma_{0\mu} \epsilon_{0-}, \nonu
\delta S^{(-)}_{0\mu\nu\rho,i}
\A = \A {\sqrt{3} \over 12} i \left( 
3 \bar\epsilon_{0+} \tau_i \gamma_{0[\mu\nu} 
\underline{\psi_{0\rho]-}} 
+ 3 \bar\epsilon_{0-} \tau_i \gamma_{0[\mu\nu} \psi_{0\rho]+} 
- \bar\epsilon_{0+} \gamma_{0\mu\nu\rho} \lambda_{0i+} \right) \nonu
\A \A - {\sqrt{3} \over 8m} i D_{0[\mu} \left( \bar\epsilon_{0+} 
\tau_i \gamma_{0\nu} \psi_{0\rho]+} \right), \nonu
\delta B_{0\mu}{}^{ij} 
\A = \A {1 \over 4} \bar\epsilon_{0+} \tau^{ij} 
\underline{\psi_{0\mu-}} 
+ {1 \over 4} \bar\epsilon_{0-} \tau^{ij} \psi_{0\mu+} 
+ {1 \over 8} \bar\epsilon_{0+} 
\tau^k \tau^{ij} \gamma_{0\mu} \lambda_{0k+}, \nonu
\delta \lambda_{0i+} 
\A = \A - 2m \tau^j \epsilon_{0+} \phi_{0ij} 
+ {1 \over 16} \left( \tau_{jk} \tau_i - {1 \over 5} \tau_i 
\tau_{jk} \right) \gamma_0^{\mu\nu} \epsilon_{0+} 
F_{0\mu\nu}{}^{jk} \nonu
\A \A - {1 \over \sqrt{3}} i m \left( \delta_i^j 
- {1 \over 5} \tau_i \tau^j \right) \gamma_0^{\mu\nu\rho} 
\epsilon_{0-} S^{(-)}_{0\mu\nu\rho,j} \nonu
\A\A - \sqrt{3} i m \left( \delta_i^j 
- {1 \over 5} \tau_i \tau^j \right) \gamma_0^{\nu\rho} 
\epsilon_{0+} \underline{S_{0\nu\rho r,j}}, \nonu
\delta \phi_{0ij} 
\A = \A {1 \over 2} \left(
\bar\epsilon_{0+} \tau_{(i} \underline{\lambda_{0j)-}} 
+ \bar\epsilon_{0-} \tau_{(i} \lambda_{0j)+} \right), 
\label{6dfsym}
\ea
where $S_{0\mu\nu\rho,i}^{(-)} = S_{0\mu\nu\rho,I=i}^{(-)}$, 
$B_{0\mu}{}^{ij} = B_{0\mu}{}^{I=iJ=j}$. 
The underlined fields are not independent fields on the boundary 
but can be expressed by the fields $\Phi_0$. 
The transformations (\ref{6dfsym}) are actually equivalent to 
those of the conformal supergravity \cite{BSvP}. 
To see the equivalence we redefine the fields as 
\ba
\tilde e_\mu{}^a \A = \A e_{0\mu}{}^a, \qquad
\tilde\psi_\mu = \psi_{0\mu+}, \nonu
\tilde T_{abc,i} 
\A = \A - 4 \sqrt{3} i m \left( S^{(-)}_{0abc,i} 
+ {\sqrt{3} \over 16m} i \bar\psi_{0[a+} \tau_i \gamma_b 
\psi_{0c]+} \right), \nonu
\tilde V_\mu{}^{ij} \A = \A -8m B_{0\mu}{}^{ij}, \qquad
\tilde\chi_i{}^\gamma = 15 m \lambda_{0i+}{}^\gamma, \qquad
\tilde D_{ij} = - 120 m^2 \phi_{0ij} 
\label{fieldredef}
\ea
and the transformation parameters as 
\be
\tilde\epsilon = \epsilon_{0+}, \qquad
\tilde\eta = 2m \epsilon_{0-}. 
\ee
By dropping tildes on the fields to avoid unnecessary complications 
in equations we obtain the transformations for fermionic 
residual symmetry as 
\ba
\delta e_\mu{}^a 
\A = \A {1 \over 2} \bar\epsilon \gamma^a \psi_\mu, \nonu
\delta \psi_\mu 
\A = \A D_\mu \epsilon 
+ {1 \over 24} \tau^i \gamma^{abc} \gamma_\mu \epsilon 
T_{abc,i} + \gamma_\mu \eta, \nonu
\delta T_{abc,i} 
\A = \A - {1 \over 32} \bar\epsilon \tau_i \gamma^{de} 
\gamma_{abc} R_{de}(Q) - {1 \over 15} \bar\epsilon 
\gamma_{abc} \chi_i, \nonu
\delta V_\mu{}^{ij} 
\A = \A - \bar\epsilon \tau^{ij} \phi_\mu 
- {1 \over 15} \bar\epsilon \tau^k \tau^{ij} \gamma_\mu \chi_k 
- \bar\eta \tau^{ij} \psi_\mu, \nonu
\delta \chi_i 
\A = \A {1 \over 4} \tau^j \epsilon D_{ij} 
- {15 \over 128} \left( \tau_{jk} \tau_i 
- {1 \over 5} \tau_i \tau_{jk} \right) \gamma^{\mu\nu} 
\epsilon R_{\mu\nu}{}^{jk}(V) \nonu
\A\A + {5 \over 32} \left( \delta_i^j - {1 \over 5} \tau_i 
\tau^j \right) \gamma^{abc} \gamma^\mu \epsilon D_\mu T_{abc,j} 
+ {5 \over 8} \left( \delta_i^j - {1 \over 5} \tau_i 
\tau^j \right) \gamma^{abc} \eta T_{abc,j}, \nonu
\delta D_{ij} 
\A = \A \bar\epsilon \tau_{(i} \gamma^\mu {\cal D}_\mu \chi_{j)} 
- 2 \bar\eta \tau_{(i} \chi_{j)} 
\label{6dfsym2},
\ea
where 
\ba
R'_{\mu\nu}(Q) \A = \A 2 D_{[\mu} \psi_{\nu]} 
+ {1 \over 12} \tau^i \gamma^{abc} \gamma_{[\mu} 
\psi_{\nu]} T_{abc,i}, \nonu
\phi_\mu \A = \A - {1 \over 16} \left( 
\gamma^{\rho\sigma} \gamma_\mu - {3 \over 5} \gamma_\mu 
\gamma^{\rho\sigma} \right) R'_{\rho\sigma}(Q), \nonu
R_{\mu\nu}(Q) \A = \A R'_{\mu\nu}(Q) 
+ 2 \gamma_{[\mu} \phi_{\nu]}, \nonu
R_{\mu\nu}{}^{ij}(V) \A = \A 2 \partial_{[\mu} V_{\nu]}{}^{ij} 
+ 2 V_{[\mu}{}^{ik} V_{\nu]k}{}^j, \nonu
{\cal D}_\mu \chi_i \A = \A D_\mu \chi_i 
- {1 \over 4} \tau^j \psi_\mu D_{ij} 
+ {15 \over 128} \left( \tau_{jk} \tau_i 
- {1 \over 5} \tau_i \tau_{jk} \right) \gamma^{\rho\sigma} 
\epsilon R_{\rho\sigma}{}^{jk}(V) \nonu
\A \A - {5 \over 32} \left( \delta_i^j - {1 \over 5} \tau_i 
\tau^j \right) \gamma^{abc} \gamma^\nu \psi_\mu 
D_\nu T_{abc,j} 
- {5 \over 8} \left( \delta_i^j - {1 \over 5} \tau_i 
\tau^j \right) \gamma^{abc} \phi_\mu T_{abc,j}. \nonu
\ea
To obtain eq.\ (\ref{6dfsym2}) we have used field equations derived 
from the Lagrangian (\ref{lag}) to express the underlined fields 
in eq.\ (\ref{6dfsym}) as 
\be
\psi_{0\mu-} = {1 \over 2m} \phi_\mu, \qquad
\lambda_{0i-} = {1 \over 20m^2} \gamma^\mu 
{\cal D}_\mu \chi_i, \qquad
S_{0\mu\nu r, i} = {1 \over 96\sqrt{3}m^2} i 
D^\rho T_{\mu\nu\rho,i}. 
\ee
The transformations (\ref{6dfsym2}) coincide with those in the 
conformal supergravity \cite{BSvP}. The transformations with the 
parameters $\epsilon$ and $\eta$ represent local super and super 
Weyl transformations in six dimensions respectively. The bosonic 
transformations of the fields defined in eq.\ (\ref{fieldredef}) 
are easily obtained from eq.\ (\ref{6dbsym}) and also coincide 
with those in the conformal supergravity. 
\par
Thus, we have shown that boundary values of the fields in the 
maximal AdS supergravity in seven dimensions form a supermultiplet 
of the $N=(2,0)$ conformal supergravity in six dimensions, and 
that local symmetry transformations in the bulk induce local 
transformations of the conformal supergravity on the boundary. 
Using these results we can compute anomalies of local symmetries 
on the boundary as in the AdS${}_3$/CFT${}_2$ case \cite{NT2} and 
obtain Ward identities for correlation functions of the boundary 
conformal field theory. Weyl anomaly in a purely gravitational 
background was obtained in ref.\ \cite{HS}. The gauge anomaly is 
easily obtained from the well-known relation between the 
Chern-Simons terms and chiral anomalies \cite{ZWZ}. 
Super Weyl anomaly can be also obtained as in ref.\ \cite{NT2}. 
\par

\bigskip

\noindent {\large{\bf Acknowledgement}}
\par
One of the authors (Y.T.) would like to thank E. Sezgin for useful 
discussions. 
This work is supported in part by the Grant-in-Aid for Scientific 
Research on Priority Area 707 ``Supersymmetry and Unified Theory 
of Elementary Particles'', Japan Ministry of Education. 
%
%
\newcommand{\NP}[1]{{\it Nucl.\ Phys.\ }{\bf #1}}
\newcommand{\PL}[1]{{\it Phys.\ Lett.\ }{\bf #1}}
\newcommand{\CMP}[1]{{\it Commun.\ Math.\ Phys.\ }{\bf #1}}
\newcommand{\MPL}[1]{{\it Mod.\ Phys.\ Lett.\ }{\bf #1}}
\newcommand{\IJMP}[1]{{\it Int.\ J. Mod.\ Phys.\ }{\bf #1}}
\newcommand{\PR}[1]{{\it Phys.\ Rev.\ }{\bf #1}}
\newcommand{\PRL}[1]{{\it Phys.\ Rev.\ Lett.\ }{\bf #1}}
\newcommand{\PTP}[1]{{\it Prog.\ Theor.\ Phys.\ }{\bf #1}}
\newcommand{\PTPS}[1]{{\it Prog.\ Theor.\ Phys.\ Suppl.\ }{\bf #1}}
\newcommand{\AP}[1]{{\it Ann.\ Phys.\ }{\bf #1}}
\newcommand{\ATMP}[1]{{\it Adv.\ Theor.\ Math.\ Phys.\ }{\bf #1}}

\begin{thebibliography}{100}
%
\bibitem{MAL} J. Maldacena, 
        The large $N$ limit of superconformal field theories 
        and supergravity, \ATMP{2} (1998) 231, hep-th/9711200. 
\bibitem{GKP} S.S. Gubser, I.R. Klebanov and A.M. Polyakov, 
        Gauge theory correlators from noncritical string theory, 
        \PL{B428} (1998) 105, hep-th/9802109. 
\bibitem{WITTEN} E. Witten, 
        Anti de Sitter space and holography, 
        \ATMP{2} (1998) 253, hep-th/9802150. 
%
\bibitem{AGMOO} O. Aharony, S.S. Gubser, J. Maldacena, H. Ooguri 
        and Y. Oz, Large $N$ field theories, string theory and 
        gravity, hep-th/9905111. 
%
\bibitem{FFZ} S. Ferrara, C. Fr\o nsdal and A. Zaffaroni, 
        On $N=8$ supergravity on ${\rm AdS}_5$ and $N=4$ 
        superconformal Yang-Mills theory, \NP{B532} (1998) 153, 
        hep-th/9802203. 
\bibitem{LT} H. Liu and A.A. Tseytlin, 
        $D=4$ super Yang-Mills, $D=5$ gauged supergravity 
        and $D=4$ conformal supergravity, \NP{B533} (1998) 88, 
        hep-th/9804083. 
%
\bibitem{NT} M. Nishimura and Y. Tanii, Supersymmetry in the 
        AdS/CFT correspondence, \PL{B446} (1999) 37, hep-th/9810148. 
\bibitem{NT2} M. Nishimura and Y. Tanii, Super Weyl anomalies 
        in the AdS/CFT correspondence, \IJMP{A14} (1999) 3731, 
        hep-th/9904010. 
%
\bibitem{BAUTIER} K. Bautier, Diffeomorphisms and Weyl 
        transformations in AdS${}_3$ gravity, hep-th/9910134. 
%
\bibitem{PPvN} M. Pernici, K. Pilch and P. van Nieuwenhuizen, 
        Gauged maximally extended supergravity in seven dimensions, 
        \PL{B143} (1984) 103. 
\bibitem{BSvP} E. Bergshoeff, E. Sezgin and A. van Proeyen, 
        (2,0) tensor multiplets and conformal supergravity in $D=6$, 
        {\it Class.\ Quant.\ Grav.\ }{\bf 16} (1999) 3193, 
        hep-th/9904085. 
\bibitem{NVvN} H. Nastase, D. Vaman and P. van Nieuwenhuizen, 
        Consistent nonlinear KK reduction of 11d supergravity 
        on AdS${}_7$ $\times$ S${}_4$ and self-duality in odd 
        dimensions, hep-th/9905075. 
\bibitem{TPvN} P.K. Townsend, K. Pilch and P. van Nieuwenhuizen, 
        Self-duality in odd dimensions, \PL{B136} (1984) 38. 
%
\bibitem{HS} M. Henningson and K. Sfetsos, Spinors and the AdS/CFT 
        correspondence, \PL{B431} (1998) 63, hep-th/9803251. 
\bibitem{AF} G.E. Arutyunov and S.A. Frolov, On the origin of 
        supergravity boundary terms in the AdS/CFT correspondence, 
        \NP{B544} (1999) 576, hep-th/9806216. 
\bibitem{ZWZ} B. Zumino, Y.-S. Wu and A. Zee, Chiral anomalies, 
        higher dimensions, and differential geometry, 
        \NP{B239} (1984) 477. 
%
\end{thebibliography}
\end{document}